\documentclass[trackchanges]{aastex701}

\begin{document}

\title{Cow-culation: Reentry Impact Risk to Livestock in the Satellite Megaconstellation Era}

\author[0000-0001-5368-386X]{Samantha~M. Lawler}
\altaffiliation{Former collaborator on Team Cow, now a collaborator on Team Goat}
\affiliation{Campion College and the Department of Physics, University of Regina, Regina, SK S4S 0A2, Canada}
\affiliation{Visiting Erskine Fellow, University of Canterbury, Private Bag 4800, Christchurch 8140, New Zealand}
\email[show]{samantha.lawler@uregina.ca}

\author[0000-0003-3257-4490]{Michele T. Bannister}
\affiliation{School of Physical and Chemical Sciences --- Te Kura Mat\={u}, University of Canterbury, Private Bag 4800, Christchurch 8140, New Zealand}
\email[show]{michele.bannister@canterbury.ac.nz}

\author[0000-0002-8974-7703]{Laura E. Revell}
\affiliation{School of Physical and Chemical Sciences --- Te Kura Mat\={u}, University of Canterbury, Private Bag 4800, Christchurch 8140, New Zealand}
\email{laura.revell@canterbury.ac.nz}

\submitjournal{Acta Prima Aprilia}

\begin{abstract}

The commercial space industry is launching more satellites into Low Earth Orbit every year.
Aotearoa New Zealand (NZ) has a thriving dairy and cattle industry.    
Unfortunately, these industries could come into (high speed) cow-llision, as the rapid launch rate and short operational lifetimes of satellites in megaconstellations like Starlink result in a high reentry rate at NZ's latitudes.
This could intersect with NZ's famously large population of livestock.  
We predict this will be an udder disaster for any cows that are hit, as they are squishy and moo-ve much more slowly than space debris.
Using a global bovine density dataset, previously published satellite casualty probability code, and a complete lack of funding to do this calculation carefully enough for submission to a peer-reviewed journal, we calculate a $\simeq 0.3$--1\% chance of a cow-sualty in NZ from reentering Starlink Gen2 debris over the next 5 years.

\end{abstract}


\section{Introduction}

Satellites are reentering Earth's atmosphere at a higher and higher rate every year.  
Operators often claim ``full demisability," i.e., complete disintegration for their reentering satellites, although this could in practice mean that potentially lethal debris, with kinetic energy greater than 15~J, occasionally reaches the ground\footnote{For instance, this piece of debris from a SpaceX Crew Dragon Cargo Trunk, which is larger in area than any of the authors, or a cow: \url{https://www.cbc.ca/news/canada/saskatoon/cp-space-junk-more-reports-1.7219927}}.
Low Earth Orbit now contains over 10,000 Starlink satellites, with over a dozen other companies scrambling to launch their own megaconstellations.
\citet{Wright2026} calculates the human casualty expectation from planned megaconstellations, finding that there is a 20\% chance of a human casualty resulting from the reentries of 30,000 Starlink satellites, assuming one piece of debris per reentry.  
This agrees well with casualty calculations from previous works \citep[e.g.,][]{Pardini2025,Hook2026}\footnote{See also these official reports from The Aerospace Corporation \url{https://csps.aerospace.org/sites/default/files/2021-08/Ailor_LgConstDisposal_20200113.pdf} and the US Federal Aviation Administration (FAA) \url{https://www.faa.gov/sites/faa.gov/files/Report_to_Congress_Reentry_Disposal_of_Satellites.pdf}}.
These other authors have completed casualty risk calculations from reentries much more carefully than us, and you should really go read their papers instead of this one!

The casualty risks extend beyond people.
Humanity is outmassed by its domesticated species, which now form the majority of Earth's total mammal biomass \citep{owid-wild-mammals-birds-biomass}. 
Foremost among the massive livestock animals are cows (\textit{Bos indicus} and \textit{B. taurus} comprise 38\%).
However, cows, like people, are inhomogeneously distributed across the planet.

Aotearoa New Zealand's legendary status as a nation of ``more sheep than people" peaked at a ratio of 22:1 in 1982, but is now at its lowest in 170 years at 4.5:1, with a national flock of 23.6 million at June 2024 \citep{Olsen2023,Birch2025}.
Cows, on the other hoof, sit at 5.8 million dairy and 3.679 million beef cattle, reflecting a long-term shift in New Zealand's agriculture. 
Globally, New Zealand (NZ) has an unusually high bovine density \citep[e.g.,][]{UNcowdata}, which is pastured outdoors. 
Additionally, NZ has a highly urbanised population ($\sim85$\%) with correspondingly low population density; large areas of the islands typically have fewer than one person per square kilometre\footnote{See for instance \url{https://en.wikipedia.org/wiki/Demographics_of_New_Zealand\#/media/File:2023_NZ_Census_Population_Density.png}}.
A person has a smaller cross-sectional area than that of a sheep or a cow$^{[citation\;needed]}$, and New Zealanders typically spend two-thirds of their time indoors \citep{Khajehzadeh02012017}. 
Clearly, the major risk in NZ from the burgeoning global space economy is to its celebrated livestock\footnote{Technically we should also check the whales, but at 2.5-4.0 km/hr \citep{noad2007}, they're a bit more of a moving target.}.

\section{Assume a spherical cow}
 
Following \citet{Wright2026}, we infer the probability of debris from a deorbiting Starlink satellite hitting bovine livestock (dairy cows and beef cattle) in NZ.
We choose Starlink because they have by far the majority of satellites currently on orbit, with over 10,000 as of this writing\footnote{See J.~McDowell's website \url{https://planet4589.org/space/con/conlist.html} for the current count.}. 
Not to mince words -- we also choose Starlink because they are the only megaconstellation thus far to have proven reentry debris hit farmland\footnote{More details here: \url{https://www.cbc.ca/news/canada/saskatchewan/2nd-piece-of-space-junk-landed-on-saskatchewan-farmland-in-2024-1.7502192}}.
Although it impacted a lentil field in Saskatchewan, Canada, not a dairy or cattle operation in NZ\footnote{Hence no mince was generated, on this occasion.}, it is an important piece of evidence that Starlink satellites do not always fully demise, at similar mid-latitudes to those we consider here. 

\begin{figure}[]
    \centering
    \includegraphics[height=3.in]{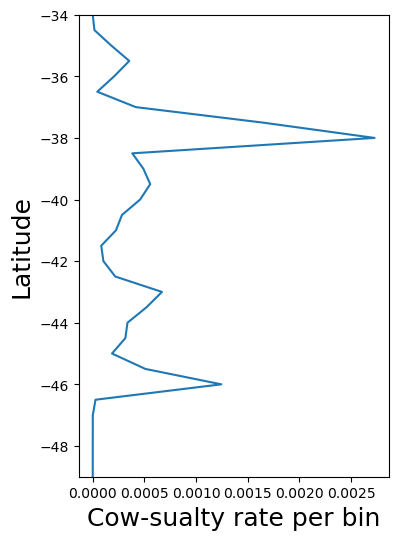}
    \includegraphics[height=3.in]{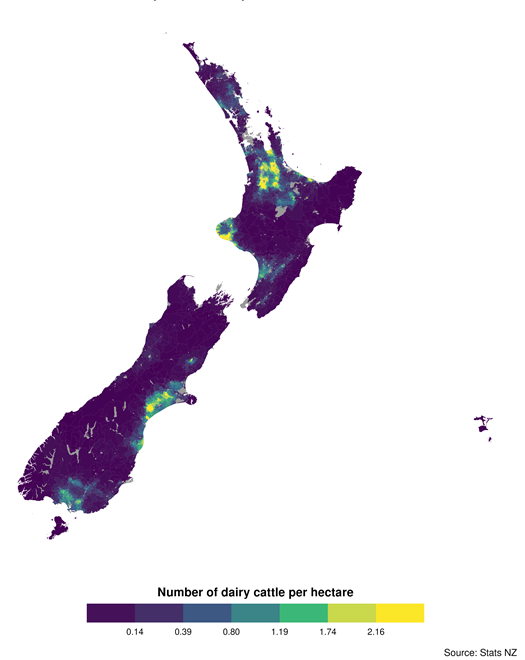}
    \includegraphics[height=3.in]{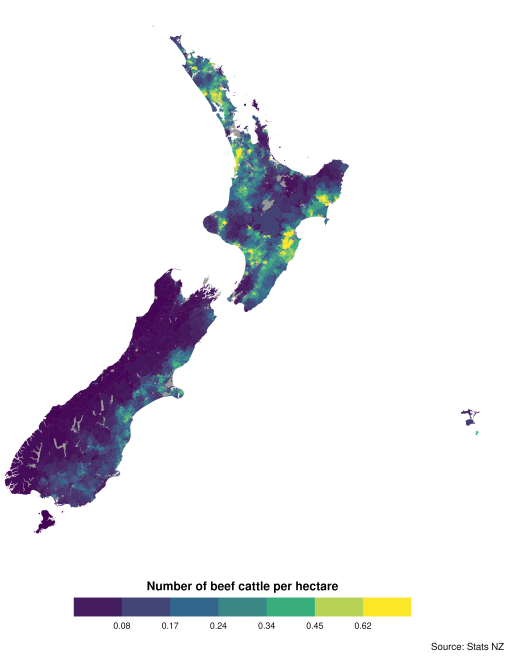}
    \caption{Left panel shows the NZ cow-sualty rate per latitude bin per Starlink Gen2 5-year cycle assuming 2m spherical cows. The centre and right panels show the 2022 livestock density per hectare of farmland, for NZ dairy cows and beef cattle respectively.} 
    \label{fig:NZcows}
\end{figure}

The spatial density of NZ's livestock is exceptionally well quantified per hectare\footnote{Statistics New Zealand run an Agricultural Production Statistics program for farms earning $>\$60$k/yr.}, as is the human population density via the Census\footnote{Although the methodology for the Census is about to change, so perhaps this will need revisiting in future work. \url{https://www.stats.govt.nz/information-releases/estimated-resident-population-2023-base-at-30-june-2023/}}.
Figure~\ref{fig:NZcows} shows the distribution of cow density, from the most recent full survey\footnote{More data available at \url{https://www.stats.govt.nz/indicators/livestock-numbers-data-to-2023/}}. 
However, one beef we had was that these data are not available from Stats NZ at this spatial resolution, only at the coarser scale of cows per NZ region \citep{livestockdensity}.
So in order to take the bull by the horns, we instead use the slightly earlier Gridded Livestock of the World \citep{UNcowdata}, summed from its native 5 arcmin ($\sim 10$km resolution) to $0.5^{\circ}$ resolution to match resolution used for human casualty estimations \citep[cf.][]{Wright2026}.

We next compute the impact probability for a megaconstellation as a function of latitude.
The orbital inclination of a satellite sets its longest dwell time above a particular latitude.
Because all megaconstellations to date use uncontrolled reentries\footnote{SpaceX claimed that they would start using ``semi-controlled reentries'' to plop Starlink satellite reentry debris into the Pacific Ocean in a document over a year ago \url{https://starlink.com/public-files/Starlink_Approach_to_Satellite_Demisability.pdf}, but the high rate of observed Starlink reentries over populated areas during the past year seems to strongly suggest they are not in fact doing this.}, the impact probability of each satellite in the model of \citet{Wright2026} is a weighted function of the time spent above each latitude.
New Zealand spans temperate latitudes across $\sim 34.5$--$47.5$$^{\circ}$S. 
As shown in Figure 2c of \citet{Wright2026}, Starlink's chosen orbits result in an increased chance of hitting a bullseye in NZ. 
The highest spike in our cow-sualty rate per latitude bin (left panel of Figure~1) is due to the orbital caustic of the Starlink 38$^{\circ}$ inclination orbital shell, which has its highest density right above Waikato and the east coast of the North Island --- prime New Zealand dairy and cattle country respectively.  There's also an obvious contribution from the intersection of the orbital caustic of the 43$^{\circ}$ inclination Starlink shell with the Canterbury dairy region, as well as the 46$^{\circ}$ shell and the Southland dairy region.
 
Megaconstellation satellites are very large, with Starlink satellite masses ranging from that of a dairy cow up to that of a full-grown bull.  
While cows are fully demisable\footnote{As tested regularly on barbecues around the motu (islands).}, satellites sometimes are not.
We follow \citet{Wright2026} and assume that each reentry results in one piece of potentially lethal debris, which could interact with the cross-sectional area of a cow. 
We calculate interaction probability using public code provided in \citet{Wright2026}\footnote{Available at \url{https://github.com/etwright1/ConstellationCasualties}}, which gives the expectation value $E_\mu$ of a NZ cow-sualty
\begin{equation}
E_\mu = \Sigma_l~\mu~p_l~w_l
\end{equation}
where we are summing over each latitude band $l$ with weightings $w_l$ and probabilities $p_l$, to account for the fly-over time and densities of the Starlink Gen2 30,000 satellite megaconstellation.
$\mu$ (mu) is, of course, the cross-sectional area of a spherical cow. (The corresponding probability of sheep impact would naturally include $\lambda$ (lamb-da) instead.) 
Cows are considered to have a zero-length mean free path; while a dairy cow walks on average 2.7 km/day to go to-and-from milking\citep{cowwalk}, each cow spends only minutes in the milking shed, and their velocity of 0.1 km/hr is negligible for this estimate.

As much as we like spherical cows \citep[e.g.,][]{Bernardinelli2021}, we did struggle with choosing an appropriate spherical cow radius\footnote{For refinement, consider the careful exploration of multipoles of the cow by \citet{lehmann2025highermultipolescow}.}, and that radius does affect the cow-culation quite strongly.  
For instance, a Holstein-Friesian cross has trunk length $1.71\pm 0.08$~m \citep{cowdimension}; the cow size distribution across the full NZ bovine population appears surprisingly less well documented\footnote{The mass distribution is characterized, but less so the size distribution.} than say, that of minor planets, but we will consider it to be flat for this work.
We find using a 2~m cow radius results in $E_\mu=0.01$, while a radius of 0.85~m results in $E_\mu=0.003$. %

Using equation 2 from \citet{Wright2026}, we can convert these expectation values to probabilities.  
Using the 2~m radius cow, this corresponds to a 1\% probability of a cow in New Zealand being hit by a Starlink Gen2 satellite over Starlink's 5 year launch-and-reentry period.  
Using a 0.85~m radius cow results in a 0.3\% cow-llision probability.

\section{No use crying over spilled milk}

It behooves us to consider the outcomes from cow-llisions.
The Outer Space Treaty and Space Liability Convention govern objects launched into orbit around the Earth.  
According to these treaties, the state where the space object was launched from is absolutely liable for damage caused on the ground in other states \citep[e.g.,][]{Leoni2026}.
For instance, NZ is the third most prolific launching state. 
NZ's Outer Space and High-altitude Activities Act (2017) and its Regulations specify that NZ has an absolute liability for damage caused by its space objects on the surface of the Earth.
Payloads launched under NZ permits\footnote{To our knowledge, no payload licences have yet been granted for the alternative case of launching a cow to space. \url{https://www.mbie.govt.nz/science-and-technology/space/apply-for-a-licence-or-permit/payloads-approved-for-launch}} have indemnity and insurance requirements assessed on a case by case basis, but if launched elsewhere, such as from the US, must be insured up to the level of the maximum probable loss \citep{cowsmack}. 
However, space debris from another state would instead be regulated by its own country's licensing and insurance requirements, which hopefully would not be all hat and no cattle.

Thus far, there have been no confirmed human or livestock deaths from reentered space debris\footnote{There is one confirmed bovine death due to a rocket that exploded, which could perhaps be classified as proto-space-debris, in Cuba in 1960: \url{https://www.nytimes.com/1960/12/05/archives/cubans-and-cows-march-in-protest-300-havana-students-mourn-beast.html}}, but due to the exponential increase in uncontrolled reentries of satellites over the past few years, casualties are getting more likely.
The steaks are high\footnote{The steaks could be potentially higher if ejected after a satellite impact, but we do not model that here.} for NZ cows: due to NZ's high bovine density and its position beneath orbital caustics of Starlink, NZ cows have relatively high risk as space debris landing sites.
Unlike some satellites, cows are fully demisable, and farmers would be very sad to lose them. 
NZ farmers will want to know how to respond if and when one of their livestock have an unfortunate interaction with a high-speed piece of space debris owned by a foreign country.
We hope the Ministry of Primary Industries, the NZ Space Agency, the Ministry of Foreign Affairs and Trade, and agricultural insurance agencies are ready and willing to help farmers who may find themselves in this situation in the future.

\section{Conclusion}

The high density of satellites in orbit is already bordering on the bull-in-a-china-shop scenario of Kessler Syndrome \citep{Kessler1978}, with various metrics \citep[e.g.,][]{lawrence2023,Thiele2025} demonstrating high stress in Low Earth Orbit.  
Some parts of New Zealand are also showing stress due to the increasingly high densities of cows and poor environmental management strategies\footnote{See for example \url{https://ehinz.ac.nz/assets/Surveillance-reports/Released_2026/Number-and-density-of-livestock-in-New-zealand_final-version-v2.pdf}}. 
Fortunately for farmers (and everyone living near dairy operations), unlike satellites in Low Earth Orbit, the cows are not moving at 7~km/s. 
While the close approach distances between cows in a given paddock are small, we can safely assume that the probability of catastrophic shattering and secondary collisions between nearby cows is effectively zero.

In New Zealand, cows are more likely to get hit by space debris than people.
Unfortunately, spherical cows do not exist in a vacuum.
This cow-culation is for one small part of the world, and also only takes into account one megaconstellation of tens of thousands of satellites.  
Companies keep asking for more and more satellites, with one million satellites filed with the ITU \citep{Falle2023}, and another million requested just a few weeks ago\footnote{Everybody seems pretty well in agreement that this is not a good idea: \url{https://au.pcmag.com/ai/116598/spacex-to-start-small-with-1-million-satellite-plan-pushes-back-on-critics}}.

Future work should ruminate on additional serious issues caused by satellite megaconstellations: light pollution \citep{Lawler2026}, atmospheric pollution \citep{Revell2026}, and increasing on-orbit collision risks \citep{Radisic2026}. 
Perhaps we should accept that we simply cannot launch so many disposable satellites into Low Earth Orbit. 
Without milking the puns dry, our livestock are cow-nting on us!

\begin{acknowledgments}
This work received no financial support, although we wish it did. 
The existence of this paper was powered by New Zealand cheese.
We thank David Hood for help understanding Stats NZ data, and Ewan Wright and Aaron Boley for their reentry casualty probability code. 
We thank our families for putting up with us writing this in our spare time and putting up with our udderly baaaaad livestock puns.
The following individuals contributed to increasing the agricultural pun density of this work: A.~Bongarzone, R.~Dorsey, J.~Forbes, M.~Hopkins,  F.\ and S.~Lawler, B.~Leicester, and E.~Wright.

Lastly, we'd like to dedicate this paper to Michael JasonSmith (1977-2 March 2026), who would have been both exasperated by the way we have cludged the cow code documentation, and delighted by the existence of a punny paper about the possible effects of poorly planned space infrastructure on NZ livestock. 
\end{acknowledgments}

\newpage

\end{document}